# AI Algorithm for Mode Classification of PCF – SPR Sensor Design

Prasunika Khare[1], Mayank Goswami[1, #]

[1]Divyadrishti Laboratory, Department of Physics, IIT Roorkee, Roorkee, India

Department of Physics, IIT Roorkee, Roorkee, India

[#]mayank.goswami@ph.iitr.ac.in

## Abstract

Photonic Crystal Fibre design based on surface plasmon resonance phenomenon (PCF-SPR) is optimized before it is fabricated for a particular application. An artificial intelligence algorithm is evaluated here to increase the ease of the simulation process for common users. COMSOL™ MultiPhysics is used. The algorithm suggests best among eight standard machine learning and one deep learning model to automatically select the desired mode, chosen visually by the experts otherwise. Total seven performance indices: namely Precision, Recall, Accuracy, F1-Score, Specificity, Matthew correlation coefficient, are utilized to make the optimal decision. Robustness towards variations in sensor geometry design is also considered as an optimal parameter.

Several PCF-SPR based Photonic sensor designs are tested, and a large range optimal (based on phase matching) design is proposed. For this design algorithm has selected Support Vector Machine (SVM) as the best option with an accuracy of 96%, F1-Score is 95.83%, and MCC of 92.30%. The average sensitivity of the proposed sensor design with respect to change in refractive index (1.37-1.41) is 5500 nm/RIU. Resolution is $2.0498 \times 10^{-5}\ RIU^{-1}$. The algorithm can be integrated into commercial software as an add-on or as a module in academic codes. The proposed novel step has saved approximately 75 minutes in the overall design process. The present work is equally applicable for mode selection of sensor other than PCF-SPR sensing geometries.

Keywords: Photonic Crystal Fibre, Surface Plasmon Resonance, Machine learning, Photonic Sensor, Mode Classification.

## 1. Introduction

Photonic Crystal Fibre (PCF-SPR) has attracted the interest of researchers working in the fields of photonics [1], biomedical [2], chemical engineering [3] [4] [5], environment monitoring [6], etc. Photonic sensors based on this design are already available in the market for several commercial applications. These sensors are small in size thus found suitable for portable applications when compared to the Kretschmann, Otto, and their hybrid configurations [7] [8]. The first paper about SPR based sensing application was reported by Liedberg *et al.* [9] in 1983. This work includes a three-layer structure consisting of glass, metal, and air. The Transverse Magnetic (TM) polarized light (p-polarization) is incident on the glass layer. It produces an evanescent field that excites electrons in the metal-dielectric (i.e., analyte) interface to reach phase-matching conditions and produces SPR. The change in the refractive index of the analyte leads to a change in the propagation constant of the Surface plasmon (SP) wave (a wave produced due to excitation of electrons in metal and dielectric interface). This SP wave is very sensitive to change in the analyte refractive index. It makes SPR a potent tool for analyte sensing applications.

PCF-SPR Sensor is shown to differentiate the analytes having similar characteristics by adjusting designing parameters [10]. The confinement loss, sensitivity, and resolution parameters of PCF-SPR can be customized by changing the following design parameters and material.

## 2.1 Effect of Material:

The analyte is a mobile molecule that gets bound to an immobile molecule located on a thin plane metal surface of the sensor [11]. SPR sensing requires a metal layer/plasmonic layer either made of gold, silver, copper, or aluminium, etc. [12]. Unlike silver, the performance of gold varies insignificantly in a hygroscopic environment. Gold, however, as a metal layer, may sometimes lead to inaccurate analyte detection as it has a *wide resonance peak* as compared to silver [13]. However, this peak broadening problem can be controlled by selecting the optimal thickness value of gold. Oxidation of silver can be suppressed by applying a coating of titanium over the silver layer. Sensor with this bimetallic material layer design has inferior performance as compared to sensor made of silver plasmonic layer [14] [15]. Silver layer with graphene coating provides better performance than pure silver [16] [17]. Aluminium, due to its higher optical loss characteristics, is not preferred [18].

## 2.2 Effect of metal layer location:

Based on metal layer location, the PCF-SPR sensor design can be broadly classified as (a) external metal layer coated and (b) internal metal layer coated sensor design. Hasani *et al.* [19] in 2006 reported an internal metal-coated design having a coating of thin gold film on the periphery of the air hole. Several internal metal film-coated liquid-filled air hole sensors were reported to improve *resolution and enhance phase matching* between core mode and Surface Plasmon Polariton (SPP) [20] [21] [22]. Selective internal coating of a thin metal film on the periphery of a nanometre size air hole is a complex process from a fabrication point of view. Therefore, to eliminate the limitation of a thin metal film, several nanowire-based PCF-SPR sensor designs are reported [23] [24] [25]. Locating the metal layer or nanowire in the vicinity of the core of sensor design increases the *loss of field*. It happens because light field intensity gets attracted toward the metal resulting in higher losses in optical fiber. At the same time, the metal layer in external metal coated sensor design is located away from the core vicinity. It decreases the loss of field in fiber while making the sensing process convenient.

## 2.3 Effect of Metal layer thickness:

Metal layer width is an important parameter as it influences the sensor's sensitivity, confinement loss, and resonance peak. As gold thickness increases, a shift (towards higher wavelength) and broadening in resonance peak is observed [26] [27]. As silver thickness increases, a shift in resonance peak is observed towards a higher wavelength, and silver shows a relatively sharper resonance peak compare to other materials [28]. The rise in silver layer thickness consistently decreases the amplitude resonance peak [29] [30]. Optimization of the thickness of metal/plasmonic layer, its location thus becomes essential.

## 2.4 Effect of Analyte location:

Placing analyte in an external layer offers the *convenience* of filling without a radial duct and offers ease of maintenance. Flow inside the external layer is expected to experience relatively more centrifugal force, thus give less reliable results if the sensor is being rotated. Rifat *et al.* [31] demonstrated that the external sensing approach reduces the fabrication complexity. The D-shape periphery of the sensor has an analyte layer in the outer surface offers easy analyte sensing [32] [13] [33].

## 2.5 Effect of Airhole location and size:

Akowuah *et al.* [34] proposed an SPR based elliptical air hole PCF sensor design. The proposed PCF-SPR sensor design with some air holes missing (explained in a later part of the article) creates leaky modes and a birefringence effect in this work. The air hole radius is decreased to enhance the coupling between core mode and Surface Plasmon Polariton (SPP) mode [28]. Junjie *et al.* [35] reported the effect of the increase in radius of air hole showed a slight decrease in amplitude of confinement loss peak and shifted toward lower wavelength. The sensitivity variation is also very less with a change in air hole radius.

## 1.1 Effect of PCF sensor shape:

The sensor design is broadly classified into three categories: 1) D Shape PCF-SPR design, 2) Circular shape external/internal PCF-SPR design, and 3) Slotted PCF-SPR design [31]. Many D-shape PCF-SPR is recently reported to have the advantage of the flow of analyte through its flat/non-convex surface [36]. However, it requires accurate polishing and etching efforts to flatten the cladding [31]. In contrast to D-shape, circular shape PCF-SPR design is relatively simple in structure. Multiple analytes can be detected by slotted sensor design.

## 1.2 Effect of pitch:

The lattice pitch (the distance between air holes) affects the performance characteristic of the sensor. With an increase in lattice pitch, the confinement loss decreases and shows shifting of resonance peak towards lower wavelength region [35]. It is because the change in pitch affects phase-matching conditions. One can refer to work by Rifat for exhaustive comparative analysis of effects of physical parameters over sensor performance [31].

## 1.3 Simulation Techniques:

Iteratively simulating the PCF-SPR design in COMSOL™ MultiPhysics (one of the several commercial software), the wavelength resonant to a range of particular analyte can be obtained by using parametric sweep over several wavelengths [37] [37][37][37]. There are several academic codes available for the same.

Machine Learning has attracted the interest of many researchers in several fields such as robotics, biomedical imaging, instrumentation design, computer vision, etc. It is also employed in the field of photonics for biosensing [38], fiber design [39], optical communication [40]. A deep learning-based approach was reported in Nanophotonic structure recognition [41] and to predict the Q factors from the displacement pattern of air holes in Nanophotonic structures [42]. The basic philosophy behind employing ML is to testing the existing ML models for photonics applications. Most of the recent works are done to predict the optical properties of Photonic Crystal Fiber [39] [43] [38]. Several ML models exist, but studies recommending the utility of a particular model for particular applications are still under development.

## 2. Motivation

Manual searching of desired modes (by selecting various involved designing parameters) for confinement loss calculation is a time-consuming process. One example is illustrated in Fig. 1. The drop-down menu of COMSOL™ MultiPhysics is shown.

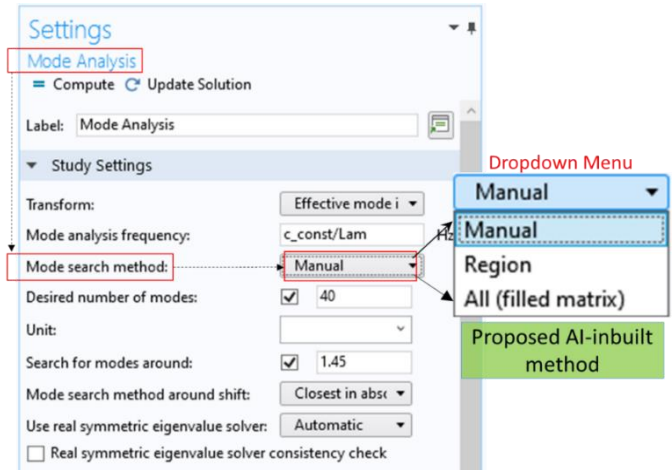

Figure 1 Proposed AI inbuilt method for mode classification.

For a given geometry, we recorded the time taken from the beginning till the desired mode was found. Manually, the user took around 55 minutes. The choice of optimal combination may vary from user to user. Basically, one does it by following the brute-force approach and visually identifying the desired field distribution/pattern in resultant images. It is a time-consuming process and requires certain experience, which may get lost if one leaves the sensor design field. A similar issue exists in other tools as well. It motivated us to propose an AI-inbuilt mode selection option. The AI-based approach is expected that the confinement loss calculation process be made easy by automatically classifying core mode (desired mode) and other modes. The secondary motivation of this paper is to carry out sensitivity analysis of several ML models for this particular Photonics sensor design application.

## 3. Theory

### 3.1 Sensor Design

PCF-SPR design has two types of modes: 1) Core guided mode and 2) Plasmonic modes. The modes propagating through the fiber's core are known as core guided mode, and these modes can be the fundamental mode(single-core), dual-core, and other modes. Plasmonic modes rein on the surface of metal-dielectric and further categorized as 1st SPP mode, and 2nd SPP mode, 3rd SPP mode, or higher-order SPP mode. These modes are shown in Figure 2. From an AI implementation perspective, desired modes are termed *class* 0.

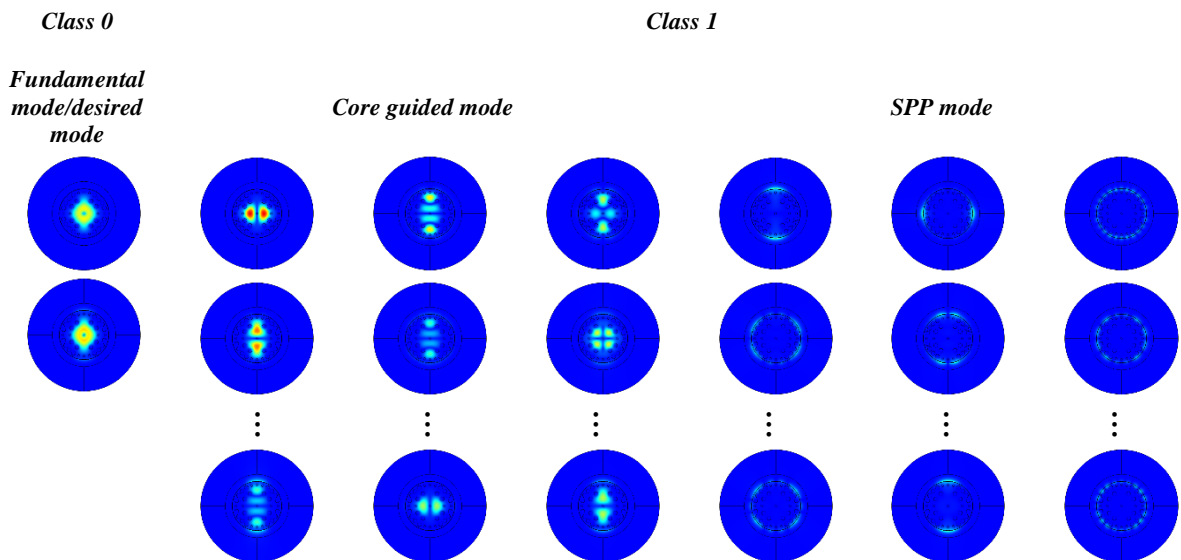

Figure 2 Field energy distribution of proposed PCF-SPR sensor.

The PCF-SPR sensor is based on the principle of the excitation of Surface plasmon by evanescent field. With the propagation of photons through the core, the evanescent field penetrates to the cladding region leads to the excitation of electrons in the metal and dielectric interface. The wave vector of evanescent field and electron on the plasmonic metal surface matches, resulting in a Surface plasmon wave. The photons propagation describing the field/energy distribution in the fiber is known as modes of the fiber.

Overall, the design parameters (pitch, the radius of air holes, the radius of core, metal width, etc.) of the sensor are tuned to excite surface plasmon in metal-dielectric interface to achieve phase match condition and optimize confinement loss. The confinement loss($\alpha$) is calculated by Eq.1.

$$\alpha \left[\frac{dB}{cm}\right] = 8.686 \times k_0 . Im(n_{eff}) \times 10^4 \qquad (1)$$

It is done by arbitrary taking some values of the parameters in the very first step, $0^{th}$ iteration. In the next step, core and SPP modes are generated by the inbuilt process in COMSOL™ MultiPhysics for a design. Further, one needs to visually search the desired field pattern out of all modes (40 modes). The fundamental mode is selected as the desired mode for our proposed design because it has the advantage of a larger travel distance with fewer losses. However, this particular choice would not affect the presented algorithm. The software also provides the value of the effective index for all modes for a range of wavelengths.

The phase match condition is achieved when the real part of the effective index (y-polarized) of one of the core modes and one of the SPP modes are equal. In the next step, confinement loss is also estimated for this case. It may not be the best of design. One can further change the design parameter repeating until the least value of the loss is obtained, $n^{th}$ iterations. The multiple rounds of varying/tuning the design parameter involve the visual search of mode more than one time. It is a time-consuming process that might vary from person to person. In the case of COMSOL™ MultiPhysics, the latter case (images of x-polarized and y-polarized) comes with arrows embedded in respective images. One has to cautiously separate those manually. In one exercise, timestamps of all steps are recorded by a user of an expert level. It took us 30 minutes to design the initial structure (parameter initialization, choosing or assigning material, building mesh, etc.). The software took 15 minutes to sample all possible modes. This time mainly depends on the wavelength range. It is found that visual searching of desired modes (y-polarized fundamental core and SPP having equal effective index) takes approximately 5 minutes. Calculation of confinement loss takes around 5 minutes. Recording of the screen (*Movie A*) for a first iteration is attached to this paper [44]. Visual mode search consumes approximately 10% (5/55 (30+15+5+5) x100) of simulation in just the first *iteration. An artificial intelligence-based approach is used to automate the process of desired mode detection problems.* Choice of AI models: 1) machine learning and 2) deep learning, however, may affect the overall outcome. Hence sensitivity analysis is necessary. A brief description is given for the photonics community.

### 3.2 Machine Learning Models

Supervised learning techniques are considered in this work because the a-prior classification of desired modes would always require users' expertise and to create training data sets. Brief descriptions of used ML classifiers with reference to sensor designing are given below.

### 3.2.1. Support Vector Machine (SVM)

The SVM, one of the preferred models, is trained using a data set $x_i \in R_n, i = 1 \ldots n.$, where $n$ is the number of data points (or the number of modes) [45]. This dataset belongs to either class1 (data excluding the desired mode) or *class* 0 (data including the desired mode), i.e., $y = +1 \; or \; y = -1$. A hyperplane ($y = wx + b$) separating these classes consist of a linear decision surface. If the set has absent surface, SVM transforms the same onto higher dimension. If $wx + b > 0$, then the pattern classified belongs to core mode(*class* 0) labelled as +1 and other mode(class1) labelled as -1. The SVM is successfully tested for various applications for example in remote sensing image classification [46], medical imaging [47], etc.

### 3.2.2. Decision Tree (DT)

Another machine learning technique used in this work is the decision tree (DT). Decision Tree architecture breaks down the problem (the mode classification in this work) into smaller solutions using splitting, stopping, and pruning [48]. The splitting of the mode classification problem is done by using the attributes (pixel values, intensity, etc.) obtained from the mode dataset. The attribute with the maximum Gini index value is taken as root nodes. Gini index is measured by using the given expression [49]:

$$Gini\ Index = \sum_{i \neq j} \sum \frac{p(C_i, T)}{|T|} * \frac{p(C_j, T)}{|T|} \qquad (2)$$

Where $i$ and $j$ represent *class* (core mode and other mode classes). |T| represent training dataset and $p(C_i, T)/|T|$ represent that the probability of the selected attribute is of class $C_i$. The $p(C_j, T)/|T|$ defines the probability that a given attribute is actually in class $j$. The nodes include information about the mode of PCF. A node can be a decision node, intermediate node, or leaf node. The decision/parent node is the node from where photonic sensor data information starts to split. The intermediate node is the chance node. A leaf node is the final outcome node (core or another mode). The branch with the highest probability (w.r.t to annotated/labeled data set) is selected as the output branch. The decision tree is very susceptible to overfitting. Overfitting occurs when the model shows low bias (less training time error) and high variance (high validation time error). The decision tree involves stopping (prevent overfitting of the model to training dataset by early stopping) and pruning (removing branches that are redundant for classification problems). The removal of a branch is performed in two steps. They are first generating many subtrees trimmed by a different amount. In the second step, they are selecting the best subtree by observing the mode classification error (True value – predicted value) produced by each subtree with an independent dataset. The early stopping and pruning reduce the overfitting of data.

### 3.2.3. Ensemble learning Techniques

Ensemble learning is a technique where many base learners are trained to solve the same problem. The base learner is also known as a weak learner because they perform slightly better than a random guess. Ensemble learning became popular because it has the ability to enhance the performance of weak learners to strong learners. A base learner consists of a base learning algorithm like a decision tree, Random Forest, Neural network, or any machine learning algorithm. Ensemble learning can consist of homogenous base learners (learner trained with one machine learning algorithm) or heterogeneous base learners (learner trained with a combination of machine learning algorithms). Generally, homogeneous base learners have preferred in ensemble-based learning method as it keeps homogeneity in outcome from each learner. The Ensemble learning-based methods used in this work are Random Forest, Gradient boost, and Adaboost. The ensemble learning methods can be categorized into 1) Bagging [50], 2) Boosting [51], and 3) Stacking [52]. In bagging, a bootstrap technique is used to subsample the photonic mode sensor training dataset with replacement. Bagging considers the most voted class (core mode or other modes) among base learners as the final predicted class. Random Forest comes under bagging. Stacking methods are not explored in this paper.

#### *3.2.3.1. Random Forest*

The bagging technique used in this work is Random Forest (RT). Breiman, in 2001, proposed the concept of randomness in trees [53]. An RF includes multiple decision tree classifiers, each provide its independent vote for $C_0 \epsilon$ core mode class or $C_1 \epsilon$ other mode class. As the random forest is made up of many Decision tree classifiers. Each tree is given a randomly subsampled input training vector obtained from the mode images dataset using the bootstrap technique. Hence, greater stability is achieved because each classifier includes variations in input data. The design needs a class selecting measure attribute to find the dissimilarity between the core mode and other modes. The preffered measure attributes are gain ratio, Gini index, Chi-square. A Random Forest mostly uses the Gini index for a measure of uncertainty of a class (core or other modes). The tree grows by learning from the training dataset and votes for either core or other mode classes. The most popular answer among all $k$ trees for mode prediction is accepted as the final answer.

*3.2.3.2. Adaboost(AB) and Gradient Boosting (GB)*

In boosting algorithm, the output of one base learner goes to another base learner sequentially to build a strong learner [54]. Schapire, in 1990 first introduced a provable boosting algorithm [55]. Later, Drucker et.al. [56] in 1993 applied the Adaboost algorithm (AB) to OCR (Optical Character Recognition ) application. The mode images obtained from simulation are given to ML models as a training dataset $X_i$, $where\ i = 1 \ldots number\ of\ modes$ and $y_i \epsilon \{0,1\}$ denotes the label core mode or other mode. Initially, Adaboost allots same weight to all core mode and other mode dataset. The base learners($h_f$) are trained iteratively by using the base learning algorithm (Decision tree in this work). At each iteration weights of incorrectly classified mode images are increased. Now, this updated weight and input $X_i$ produce another base learner by again using the base learning algorithm. This process is repeated $f$ number of time and the final predicted mode of SPR-PCF sensor is decided by majority vote from each base learner. Another boosting algorithm used in this work is Gradient Boosting**.** The only difference in Adaboost and gradient boost algorithm learning process is that GB uses a negative gradient of the loss function to minimize the error at each learning round, whereas Adaboost minimizes error by weight update.

*3.2.4. Naive Bayes(NB)*

A probabilistic-based ML model known as Naive Bayes (NB) is applied in this paper. It is based on the Bayes theorem with an assumption that the attributes of a given class(core mode or other mode class) are independent [57] [58]. This assumption is known as the Bayesian assumption, and the classifier involving this Bayes assumption is known as the Naive Bayes classifier.

*3.2.5. Stochastic Gradient Descent(SGD)*

Stochastic gradient descent belongs to the modern ML technique. It is based on optimization of a given function with gradient-based learning and with decreasing step size. Duchi et al. in 2011 investigated the various adaptive step size. Bach and Maulines in 2011 reported constant batch size with random sub-sampling to optimize the result of gradient descent algorithm.

### 3.3 Deep learning-based method

The convolutional neural network (CNN) is also tested in this paper for the mode detection problem. CNN model has gained immense popularity in Image classification problems due to its accurate prediction, but it has a disadvantage of large computational time and requires considerable annotated data in the training phase [59].

Besides having several choices of models, more than one performance indices may be required to evaluate their sensitivity towards data in hand. It is briefly described below:

### 3.4. Performance Index

In this paper, we have used seven indices to evaluate the ML model capability. Each performance index provides specific information. The performance evaluation parameter depends on the elements of the confusion matrix shown in Table1.

| | | Table 1 Confusion Matrix | |
|---|---|---|---|
| | | Predicted Mode | |
| | | Class 0 (desired mode) | Class 1 (Other modes) |
| Actual Mode | Class 0 (desired Mode) | **True Positive** | **False Negative** |

|  | Class 1 (Other Mode) | **False Positive** | **True Negative** |

True Positive($t_p$): number of modes that are actually the desired mode and also predicted as the desired mode, True Negative($t_n$): number of modes that are the other modes and also predicted as other modes, False Positive($f_p$): number of modes which are other modes but predicted as the desired mode, False Negative($f_n$): number of modes that are desired mode but predicted as other modes. The important factor to be considered while evaluating the classification model's performance is the class imbalance problem. The class imbalance problem is defined as when the number of data points in a class is far less or more than another class. Following are the definition of performance indices with brief description [60]:

*3.4.1. Precision:* Precision is the ratio between the number of core mode images that are correctly classified (as core mode images by the code, $t_p$) to the number of images that are classified as core mode ($t_p + f_p$):

$$\text{Precision} = \frac{t_p}{t_p + f_p} \quad (2)$$

This index is biased towards one class as its calculation includes only $t_p$ in its numerator.

*3.4.2. Recall:* Recall is the ratio between the number of core mode images correctly classified ($t_p$) to the number of true core mode images actually exist (had it been classified manually, $t_p + f_n$):

$$\text{Recall} = \frac{t_p}{t_p + f_n} \quad (3)$$

The Recall is also a biased form of measuring parameter because it also includes $t_p$ in its numerator.

*3.2.3. Specificity:* Specificity describes the classifier's ability to classify negative (other modes) class correctly. It is the ratio of correctly classified other modes to the actual number of other mode.

$$\text{Specificity} = \frac{t_n}{t_n + fp} \quad (4)$$

Specificity is a biased form of measure as it gives information about only negative class (other modes) classification outcomes.

*3.3.4. Accuracy:* It is the ratio of the correctly classified modes (core and other modes) to all of the modes.

$$\text{Accuracy} = \frac{t_p + t_n}{(t_p + t_n + f_p + f_n)} \quad (5)$$

In case of imbalance dataset accuracy provides an unreliably high measure even if minority class is incorrectly predicted. It happens because in the case of class imbalance, minority class datasets ($t_p$) << majority class ($t_n$). So, if $t_n$ is large, then the accuracy also becomes large [61] [62].

*3.3.5. $F_\beta$ Score:* $F_\beta$ Score is the weighted harmonic mean. It includes $t_p$, $f_n$ and $f_p$ in its calculation.

$$F_\beta \text{ score} = \frac{(1+\beta^2) \times t_p}{(1+\beta^2) \times t_p + \beta^2 \times f_p + f_n} \quad (6a)$$

$$F_1\text{-Score} = \frac{t_p + t_p}{t_p + t_p + f_p + f_n} \quad (6b)$$

We consider $\beta=1$ to balance the weightage of Recall and Precision equally in this work. The $F_1$-Score has the property of showing variation on class swapping [63]. As compared to accuracy (Eq.4), the F1 score does not include a negative class element($t_n$). If all the data elements of a particular class (suppose class 1) gets incorrectly classified, then F1-score become 0 even if another class (class 0) data elements are classified correctly. The disadvantage of F1-score makes it a biased measuring parameter [64]. Nevertheless, F1-score is the widely used performance index in machine learning.

*3.3.6. Matthews correlation coefficient(MCC):* The MCC measure was first introduced by B.W Mathew in 1975 [64]. MCC is high only if the prediction obtained is correct in all confusion matrix parameters and overcomes class imbalance problems. Therefore, it is a more reliable measure parameter. It varies between -1 to 1.

$$\text{Matthews correlation coefficient (MCC)} = \frac{(t_n \times t_p - f_n \times f_p)}{\sqrt{(t_p+f_p) \times (t_p+f_n) \times (t_n+f_p) \times (t_n+f_n)}}. \qquad (7)$$

In case a particular model classifies both the fundamental modes(minority class) and other modes(majority class) incorrectly $tn \times tp - fn \times fp < 0$, thus resulting value of MCC<0 and vice-versa. The denominator part ensures that the MCC also reflects how severely the model is giving incorrect classification. The MCC failed to perform well in some reported cases like *albelt extreme*[65]. If Tp=0, fn=2, Tn =38, are fp=0, just one of the few worst-case scenarios (here for imbalanced data set) result Precision =0/0, Recall =0, Specificity=1, Accuracy=~1, F1 score=0, MCC=0/0, highlights that none of these indices are stand-alone, good enough to estimate the performance of a model. Standard libraries take apt measures in case of undefined values.

# 4.Methodology

The flowchart of this work is illustrated in figure 2. The section is described into three main subsections **Step1:** Simulation: Photonic Sensor modelling, **Step2:** AI-based mode detection and evaluation of models, **Step 3:** Confinement loss, Sensitivity, Resolution calculation.

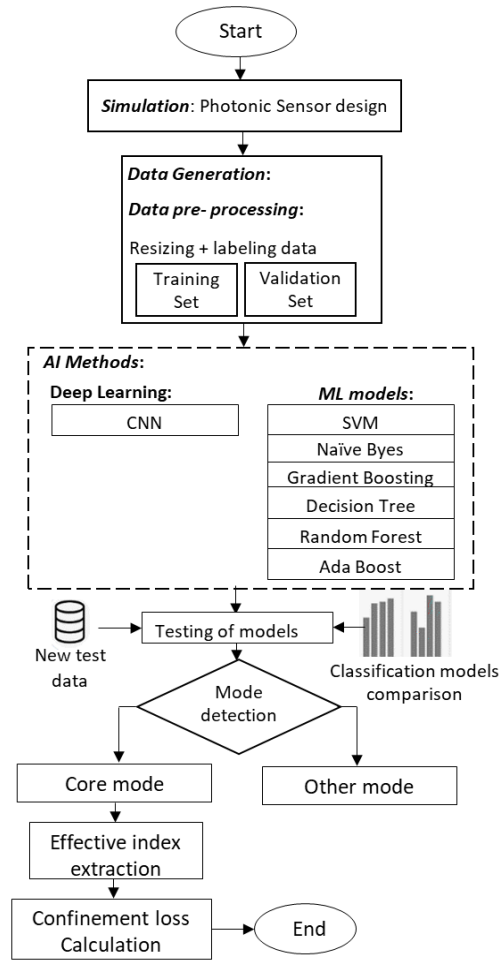

Figure 3 Flow chart.

The First step in this flowchart is to simulate PCF-SPR sensor design in COMSOL™ MultiPhysics software. The second step is to generate and pre-process the traveling mode dataset. The training and validation dataset is prepared to train and validate the various machine learning and deep learning models. These models are trained to classify fundamental modes and other modes. The seven performance measuring index are used to measure the ability of classification of models to classify modes for new test datasets. After classification, the desired mode (fundamental mode) effective index is extracted for the calculation of confinement loss. It automates the processes of manual detection of mode. The following section explains the methodology now.

## 4.1 Photonic Sensor Modelling

The proposed PCF-SPR design structure and materials are selected by considering all of the loss and sensitivity affecting parameters mentioned in the introduction section. The experimental setup of the PCF-based sensor design is illustrated in Figure 4(a). The light source used is a LASER, an inlet-outlet section is present for analyte filling. The parameters of the photonic sensor are varied. Pitch $p_1$ and $p_2$ are changed between $1.0 - 2.0\ \mu m$ and $0.5 - 2.0\ \mu m$ respectively. The radius of the core ($d_c$) is $0.10 - 0.2\ \mu m$, radius of the small hole ($d_1$) is kept equal to $d_c$, the radius of a larger hole ($d_2$) is varied between $0.1 - 0.35\ \mu m$, gold layer width is varied from $30 - 60\ nm$, analyte layer width is varied from $0.8 - 0.9\ \mu m$, and Phase Matched Layer(PML) width is varied $3.0 - 3.5\ \mu m$. A shift in resonance peak can be observed with a change in the analyte. We noted the change in confinement loss and sharpness of peak with change in the gold layer thickness ($30, 40, 50, 60\ nm$) to avoid broadening of the resonance peak. The element size of the mesh kept extra fine. The parametric sweep of wavelength is in the range $0.6 - 0.9$ µ$m$ in COMSOL$^{TM}$. The number of modes is set using the mode analysis window. We performed a simulation on five values of the number of modes: $20, 40, 60, 80$, and $100$. In this paper, the modes are searched near the refractive index of 1.45. The material used in the proposed photonic sensor design is fused silica, gold, analyte (depends on user), air. As a result, several designs were evolved during the successive *iteration* of the simulation. Some of the intermediary designs obtained during these *iterations* are shown in figure 3(b). The respective plots (to optimize the confinement loss) are also shown.

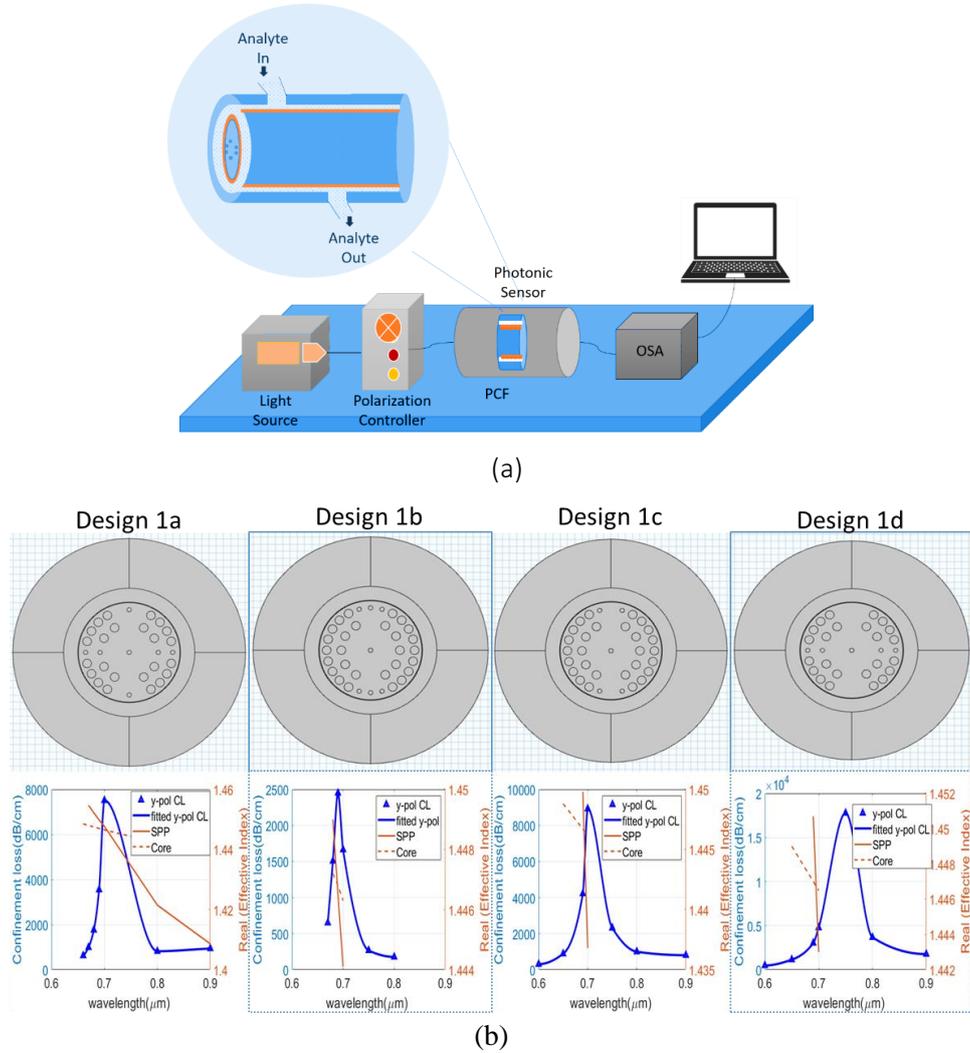

Figure 4 (a) Graphical Illustration of Experimental Setup (b) Few of the successive designs of PCF-SPR sensor.

The refractive index of fused silica material is given by the seller equation[66]. The dielectric constant of gold is provided by the Drude-Lorentz model equation[66]. The analytical relations of confinement loss, wavelength sensitivity, and the resolution of the proposed sensor are taken from a previous work obtained [66].

### 4.2 AI model implementation

Total seven machine learning models are used: (1) Support Vector Machine(SVM), (2) Random Forest(RF), (3) Decision Tree(DT), (4) Naive Byes(NB), (5) Gradient Boosting (GB), (6) Stochastic Gradient descent(SGD), and (7) Adaboost(AB).

We used the Keras library (open source neural network library version 2.1.5) to set up the CNN framework for photonic mode classification. The PCF-SPR mode dataset obtained from COMSOL$^{TM}$ software is fed as input to the deep learning CNN model. The model is trained to extract the relevant features from input images. These input images are pre-processed by resizing them to minimize the model computation cost. The following steps are involved in using Deep learning models:

### 4.2.1. Data & Image Pre-processing:

The dataset generated is an imbalance in nature. The sampling technique is applied to balance the data in each class. The most common techniques are oversampling and undersampling. Under-sampling is preferred where the dataset in the minority class is sufficiently large. In the dataset, the minority class dataset contains only two images (X-polarized core mode and y-polarized core mode). Oversampling technique (random duplicating the dataset from minority class to balance the probability of classification) is preferred. Although duplicating the dataset may over-fit the results, but in this dataset, physical characteristics variation is insignificant[67].

### 4.2.2. Model Training and output:

We compared the performance of the above seven models with respect to change in: (1) number of modes, (2) sensor geometry design. The ML model is trained with datasets having two class distributions. C$lass$ 0 includes two images: (1) x-polarized fundamental mode, (2) y-polarized fundamental mode, and $class$ 1 includes ($m - 2$) other mode images. Where $m$ is the total number of modes. Training dataset includes $m = 10, 20, 40, 60, 80$, and $100$ modes sets. Each mode dataset is generated at a particular analyte refractive index of 1.38. The testing set includes a total of 50 images. It contains 50% fundamental mode images and 50% other modes. The testing dataset is generated using an analyte of refractive index between $1.37 - 1.4$. Inbuilt machine learning models are used from the Scikit Learn library (version 0.16.2). Inbuilt parameters and definition of the models kept as it is to offer reproducibility in results. Except in one modeling instance, the controlling parameter of the depth of the decision tree (referred to as "$max\_depth$") is tuned equal to 2.

The deep learning model is constructed using the Keras framework. It employs TensorFlow as its backend. *Python (version 3.7.7)* is used as a programming language. The conventional CNN model information is given in Table 2.

*Table 2 CNN layer structure*

| S.No. | Type | Channel of filters | Filter/Pool size | Activation |
|---|---|---|---|---|
| 1 | Convolution Layer | 32 | 3,3 | ReLU |
| 2 | MaxPooling Layer | -- | 2,2 | |
| 3 | Dropout | 0.2 | -- | |
| 4 | Convolution Layer | 64 | 3,3 | ReLU |
| 5 | MaxPooling Layer | - | 2,2 | |
| 6 | Dropout | 0.2 | -- | |
| 7 | Dense Layer | 128 | -- | ReLU |
| 8 | Dropout | 0.2 | -- | -- |
| 9 | Dense Layer | 2 | -- | Sigmoid |

The convolution layer 2D layer helps in extracting the features from the given mode training dataset. Batch normalization is performed to allow a higher learning rate and to normalize the training processes. The max-pooling layer reduces the dimension of the feature map and computation cost. In this work, the pooling size is set to (2,2). The activation function used is Rectified Linear Unit (ReLU). Dropout of 0.2 is introduced to prevent the CNN model from overfitting. The model used Adam optimizer and categorical cross-entropy as loss functions. The CNN model is fit to the training dataset with a batch size of 64. These two parameters were optimized. In an imbalanced dataset case, if we decrease batch size below the number of the training dataset (in our case 40) then the model may not get trained to minority class data. So, we considered a batch size of 64. The number of the epoch is varied from 10 to 100. We have used epoch equals to 50 for fast processing and by observing the saturation level. HP Z6 workstation having Intel® Xeon 3106 processor with 128 GB RAM, Nvidia Quadro P1000 4 GB

GDDR5 GPU is used to execute the codes (written in Python). The codes with a readme file, trained weights, and sample data are provided with this paper.

## 5. Results

### 5.1 Final Design of PCF-SPR

After several designing iterations, the finally evolved design is shown in Figure 5(a). It represents the proposed PCF-SPR sensor design. It is a two-layer circular lattice PCF design. Two circular shape air holes ($90^o$ and $270^o$) anti-clockwise are excluded deliberately in the inner ring to enhance the evanescent field. On top/bottom of these, in the outer ring, three consecutive small radius air holes are included. It is to create a birefringence effect and to obtain a sharp confinement loss peak. Design 1a in Fig. 4(b) shows the basic (very first) structure. The achieved optimized parameters are listed in Table 3.

*Table 3 Sensor design parameter specification*

| Parameter Specification | Values ($\mu m$) |
|---|---|
| Radius of core ($d_c$) | 0.15 |
| Radius of small holes ($d_1$) | 0.15 |
| Radius of large holes ($d_2$) | 0.30 |
| Gold layer width | 0.04 |
| Pitch ($p_1$) | 2.0 |
| Pitch ($p_2$) | 1.0 |
| PML layer width | 3.5 |
| Analyte layer width | 0.98 |

We note that anatomically these two images (initial and final) has a difference of four holes ($80^0, 100^0, 260^0$ and $280^0$ or simply top and bottom 2 each). These extra holes in the outer ring, however, brought the average confinement loss is of an order of 1000. Figure 5(b) shows the existence of SPR and the distribution of field intensity. A similar distribution is observed in intermediatory designs; however, the phase matching (intersection of SPP and Core mode with overlapping of peak (21.06 dB/cm at $\lambda = 0.69 \mu m$) as shown in Fig. 5(c). The phase-matching condition is visibly getting validated as SPP and core both modes are traveling simultaneously (Fig. 5(b)). The location where SPR is generated is used for analyte sensing.

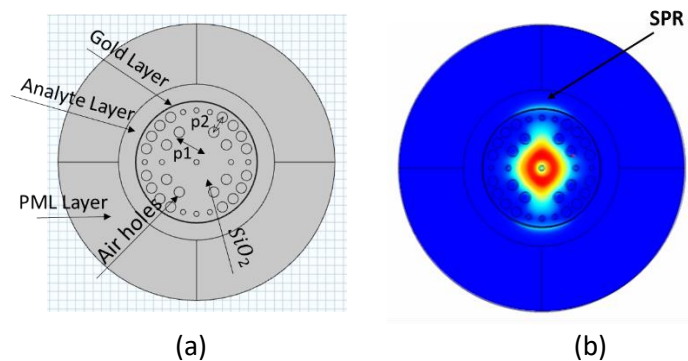

(a)        (b)

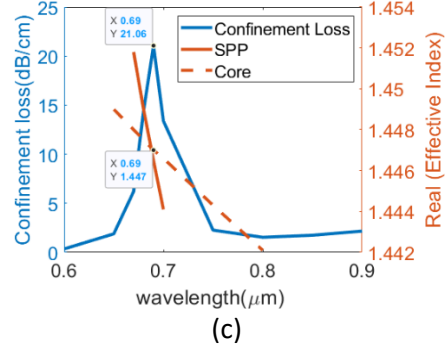

(c)

Figure 5(a) Proposed PCF-SPR sensor design, (b)SPR mode, (c) Confinement loss curve at analyte refractive index of 1.38.

Figure 6(a) shows the change in Confinement Loss(CL) with a change in the analyte. As the analyte refractive index varies, the effective mode index changes, and it leads to a change in resonance wavelength. A shift in resonance wavelength is observed towards a higher wavelength with an increase in refractive 1.36 to 1.41. The performance sensor is analyzed here using the wavelength interrogation method, as shown in Fig. 6(b). The wavelength sensitivity ($S_\lambda = |\Delta\lambda_{peak}/\Delta n_a|$) is defined as the change in resonance wavelength with a change in analyte refractive index [66]. From Fig. 6(b), one can evaluate that the average wavelength sensitivity of this design is 5500nm/RIU.

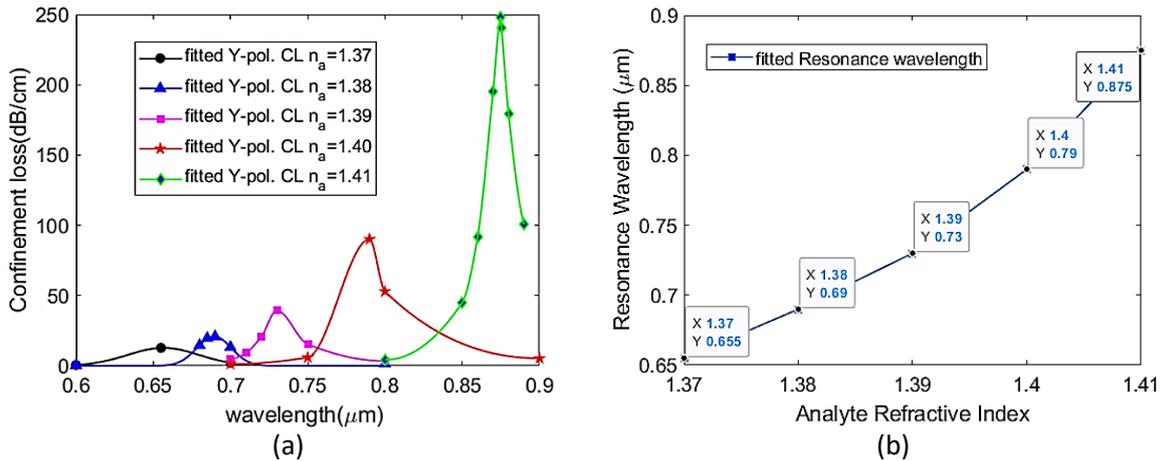

Figure 6 a) Confinement Loss variation with change in analyte and b)Resonance wavelength change with analyte refractive index.

The maximum sensitivity observed at refractive index 1.4 is $8500 nm/\text{RIU}$, and the minimum sensitivity at a refractive index of 1.37 is $3500 nm/\text{RIU}$ considering the resolution of the spectrometer as $0.1 nm$. The average resolution of the proposed sensor obtained using $R = \Delta n_a \times \Delta\lambda_{min}/\Delta\lambda_{peak}$ is $2.0498\times 10^{-5}$ RIU$^{-1}$. The proposed sensor has greater wavelength sensitivity compared to many previously reported sensors shown in Table 4. Few of the reported sensors have shown better performance as compared to ours[68][29].

*Table 4 Comparison of the proposed design with other reported structures*

| PCF-SPR sensor design | Refractive Index detection Range | Wavelength (μm) | Sensitivity (nm/RIU) | Ref. |
|---|---|---|---|---|
| HC-PCF with side channel | 1.33-1.37 | 1.40-1.70 | 1145.0 | [69] |
| PCF without core | 1.34-1.40 | 1.53-1.50 | 2278.0 | [70] |
| SM-PCF with sides polished | 1.34-1.40 | 1.48-1.46 | 4365.5 | [71] |
| Proposed SM-PCF with circular plasmonic layer design | 1.37-1.41 | 0.60-0.90 | 5500.0 | |

The application (mainly cancer cell-based assays) of the proposed design (w.r.t range of refractive index 1.37-141) one can refer to previously reported work [2][72].

## 5.2 AI-based mode detection

In this section, sensitivity analysis results of AI-based models are discussed. AI-based models are trained using $m = 20, 40, 60, 80$ and $100$ modes. Modes greater than $100$ are not considered because with the higher number of modes, changes in field patterns are negligible. This training, validation, and testing dataset are generated by COMSOL$^{TM}$ software. Fig. 7 shows plots of all performance indices vs. the number of modes $m$. According to the definition (Eqs. 2-7), a model would be best if the value of Precision, Recall, Specificity, accuracy, F1-Score, MCC would remain unaffected (and unit in magnitude) as the number of modes increases ($t_n$ is increased). It is shown that Precision and Specificity remain close to the unit except for initial cases of CNN and SGD models. It confirms that Precision and Specificity are biased towards one class. SGD has the highest (but with minor variation) recall value. SGD has identified minority class (core mode) well but under-performs to classify the majority class (other modes) as far as F1-Score and MCC values are concerned. SVM shows the least variance (w.r.t modes) and attains the highest value (unit) except for Recall. CNN has the second-highest values for all 6 performance indices but has the highest variation for a low number of modes. However, as the imbalance increases in the data set, the performance of CNN gets poor. We note that oversampling is performed in all our cases. As the number of total modes increases, the oversampling adds more number of core modes. For example, for 20 modes data set has 2 fundamental modes and 18 other modes. Oversampling would add 16 fundamental modes to achieve equality to 18 other modes in the training set. If the total number of modes is 100 then oversampling would add 96 core modes. It infers that CNN is performing well as compared to other methods (except SVM) when oversampling has the least effect, i.e., most natural modeling. Naive Bayes performed poorly (Recall of 40%) with identifying minority class data. The DT, RF, AB showed similar performance in the recall plot. Specificity is mostly high because most of the models performed well in classifying majority class data. SGD has the lowest Specificity.

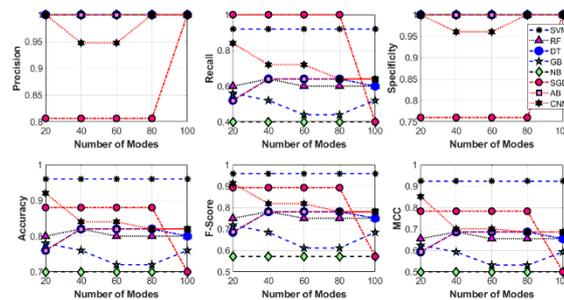

Figure 7 Performance assessment of AI based model with increase in number of modes. Increase in number of modes increase imbalance of core mode and other modes in training data. CNN and SGD showed minimum precision and specificity compared to all other models. SVM and SGD performed better in terms Recall. SVM showed higher accuracy, F-score and MCC.

Precision boxplot shows CNN and SGD showed a larger standard deviation in all performance indices (Fig.8). The mean value of Recall for NB is the least compared to other models. The SVM has the highest Accuracy-Score and MCC measure when compared to other AI-based models. SVM showed the best performance with an average Precision is 100%, Recall of 92%, Specificity is 100%, an average accuracy of 96%, F1-Score is 95.83%, and MCC of 92.30%. At $m = 100$ the Adaboost showed 4% better F1-Score compared to Decision Tree showed an average F1-Score of 75.5%, and MCC is 66.1%. Random forest(RF) shows a slight decrease in accuracy with an increase in the number of modes compared with the decision tree. Gradient boosting showed F-score 66.17%, and MCC is 57.43%. So, in the PCF-SPR dataset, the SVM performed best among all models.

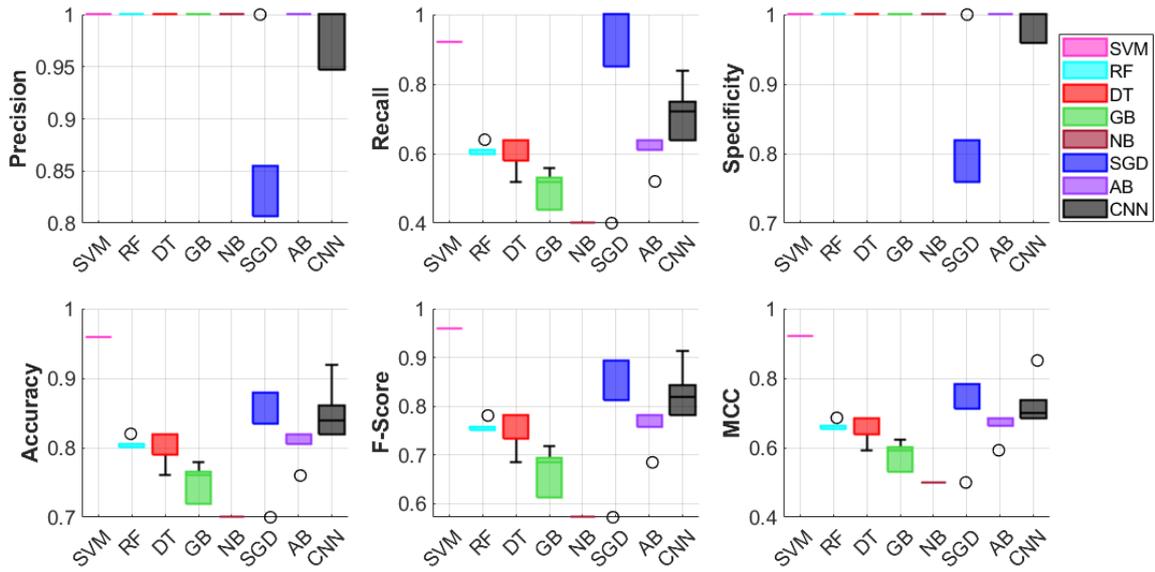

Figure 8 models performance, a) Precision, b) Recall, c) Specificity, d) Accuracy, e) F-Score, and f) MCC.

The robustness of the AI model with respect to a slight change in geometry/design of PCF-SPR is analyzed and reported in Fig. 9. We trained the AI models with the proposed geometry (geometry 1) field pattern dataset and tested the trained model with geometry 1 as well as on geometry 2 field pattern images.

The proposed PCF-SPR design (Fig. 5(b)) is denoted as geometry 1, and Fig. 2(1d) geometry is denoted by geometry 2. Fig.9 shows that the SVM has the highest MCC as well as robustness to change in geometry. Gradient boosting performed well in geometry 2 classification but poor in geometry 1. CNN and SVM showed less variance with changes in geometry. For NB, geometry 2 encounter 0/0 problem and is normalized to zero.

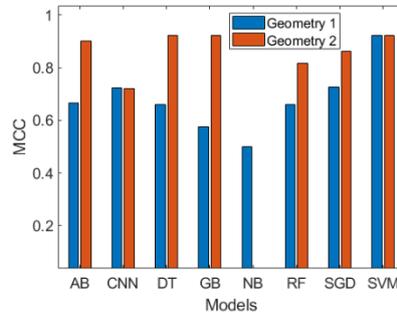

Figure 9 Comparison of Model robustness with change in PCF-SPR sensor design. The AI- models are trained using geometry 1 design mode field pattern images at a refractive index of analyte-1.38 and tested on a new dataset obtained from two geometry designs 1) Geometry 1 is the proposed design (figure 5 (a)), 2) Geometry 2 shown in figure 4 (a)- design-1d.

We ran the training codes using same data set, multiple times, testing which model would experience "reproducibility crisis" (Hutson, 2018; Raff, 2019). The codes are executed 5 times to evaluate which of the AI model would suffer the "reproducibility crisis"[73], [74]. Figure 10(a) shows the Standard deviation ($\sigma_{5\ Trials}$) of performance indices for all 8 AI models when 100 modes were used. CNN experiences the crisis most as its standard deviation remains maximum even in the case of classification of other modes as well. SVM, GB, NB, and AB show zero variation in results. Random forest performs better than stochastic gradient descent method. Precision and Specificity fail to exhibit any variation except for CNN. Only, F1-Score reflects minor difference in performance evaluation between RF and SGD. Figure 10(b) shows variation of models w.r.t number of modes using MCC.

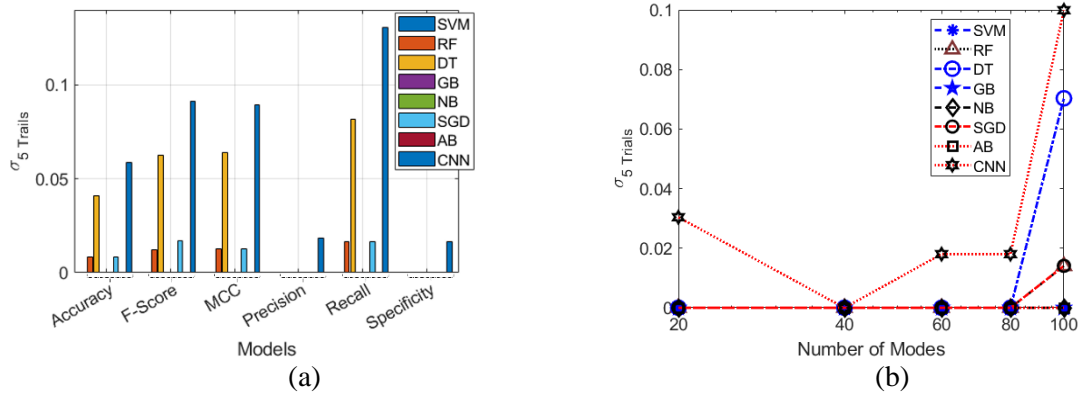

(a)                                                                                      (b)

Figure 10. Robustness of models as per reproducibility criteria, a) shows standard deviation (for five trails using 100 modes) vs. model plot for all performance indices: CNN is least reliable as far reproducibility is a concern, and b) shows standard deviation plot for (5 trails) when the number of modes is varied.

Fig.11 represents the final mode classification results obtained by using an AI-based approach. Fig. 5(a) shows the figures which appear in the graphic window of COMSOL $^{TM}$ after computation of effective index and traveling modes field patterns. Fig. 5(b)-(f) are generated during the simulation at a different number of mode analyses. The AI-based approach classifies the fundamental mode (*class* 0) and other modes (*class* 1) along with its effective mode index. This associated effective mode obtained is then used for confinement loss calculation. However, this work can be extended from binary class to multiclass classification problem by following the same AI algorithm to classify SPP modes (class 2) along with fundamental modes (*class* 0) and other modes (*class* 1).

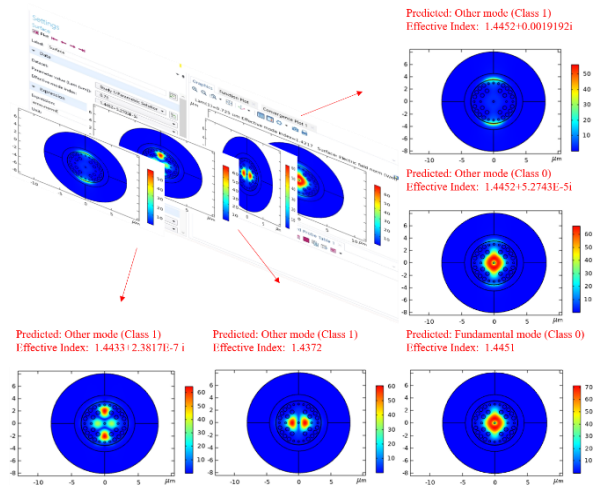

Figure 11 a)Generation and visualization of modes using COMSOL$^{TM}$ Graphic window representing the traveling mode field patterns in sensor design at analyte refractive index of 1.38 (Example 40 modes generated include 40 different field pattern images ), b) represents the results of trained model(SVM), the predicted class of 1$^{st}$ SPP mode using trained AI-based models is class 1 or other mode, and its associated effective index is obtained from simulation, c) predicted mode is class 0 or fundamental mode, d) predicted class is class 0, e) predicted class: class 1 or another mode, and f) predicted class: class 1 or other class.

## 6. Conclusion

AI algorithm is presented for automatic mode classification, a necessary and visually laborious step in PCF-SPR designing. Sensitivity analysis is performed to choose an optimal AI model to avoid the effects of the imbalanced nature of data. The aptness of performance indices is also studied. A Novel PCF-SPR design is proposed using this algorithm. Following are the pointwise conclusion:

1. The SVM is the most reliable and robust method for this dataset. GB and NB are the second best choices.
2. Manually, it took us 825 minutes to come up with the best possible design using the proposed configuration. The same process was repeated using the proposed algorithm and was completed in under 750 minutes. The proposed algorithm helps to save 75 minutes in this case.
3. Specificity and Precision showed biased behavior towards one class. The MCC (except the undefined cases) and F-Score parameter index would be a better choice to analyze the performance of the classification model.
4. CNN is the most unreliable method for this classification as after 100 modes, and images fail to provide enough structural variations.

The presented ML approach classifies single-core mode; however, it can be used to classify other modes, including 1$^{st}$ SPP mode, by converting it to a multiclass problem. *We recommend that the proposed algorithm with all AI methods must be embedded in simulation software /codes.*

### Acknowledgement:
We would like to acknowledge Prof. Sachin K. Srivastava for inspiring Ms. Prasunika to initiate this work. This work is partially supported by GRANT Code I.M.P./2018/001045 by IMPRINT-II by S.E.R.B., Government of India.

### CRediT authorship contribution statement
Mayank Goswami: Methodology, Investigation, Writing, Visualization, Supervision, Funding acquisition. Prasunika Khare: Conceptualization, Software, and Methodology.